\documentclass{article}

\usepackage[numbers]{natbib}         
\usepackage[colorlinks]{hyperref}    
\usepackage[english]{babel} 
\usepackage{amssymb}
\usepackage{amsmath}
\usepackage{txfonts}
\usepackage{mathdots}
\usepackage[classicReIm]{kpfonts}
\usepackage[dvips]{graphicx} 
\usepackage[a4paper, portrait, margin=1in]{geometry}
\usepackage{tabularx}
\usepackage{longtable}
\usepackage{multirow}
\usepackage{booktabs}
\usepackage[labelsep=period]{caption}
\usepackage{makecell}
\usepackage{arydshln}
\begin{document}
	\setlength{\parindent}{0pt}
	\setlength{\parskip}{1ex}
	
	\textbf{\Large Patient-Specific CBCT Synthesis for Real-time Tumor Tracking in Surface-guided Radiotherapy}
	
	\bigbreak

	Shaoyan Pan$^{1,2}$, Vanessa Su$^{2}$, Junbo Peng$^{1}$, Junyuan Li$^{3}$, Yuan Gao$^{1}$, Chih-Wei Chang$^{1}$, 
	Tonghe Wang$^{4}$, Zhen Tian$^{5}$ and Xiaofeng Yang$^{1,2*}$

	1Department of Radiation Oncology and Winship Cancer Institute, Emory University, Atlanta, GA 30322, USA
	
	2Department of Computer Science and Informatics, Emory University, Atlanta, GA 30322, USA
	
	3Department of Biomedical Engineering, Johns Hopkins University, Baltimore, MD, 21205
	
	4Department of Medical Physics, Memorial Sloan Kettering Cancer Center, New York, NY, 10065
	
	5Department of Radiation and Cellular Oncology, University of Chicago, Chicago, IL, 60637

	\bigbreak
	\bigbreak
	\bigbreak

	\textbf{*Corresponding author: }
	
	Xiaofeng Yang, PhD
	
	Department of Radiation Oncology
	
	Emory University School of Medicine
	
	1365 Clifton Road NE
	
	Atlanta, GA 30322
	
	E-mail: xiaofeng.yang@emory.edu

	\bigbreak
	\bigbreak
	\bigbreak
	\bigbreak
	\bigbreak
	\bigbreak

	\textbf{Abstract}

	In this work, we present a new imaging system to support real-time tumor tracking for surface-guided radiotherapy (SGRT).  SGRT uses optical surface imaging (OSI) to acquire real-time surface topography images of the patient on the treatment couch. This serves as a surrogate for intra-fractional tumor motion tracking to guide radiation delivery. However, OSI cannot visualize internal anatomy, leading to motion tracking uncertainties for internal tumors, as body surface motion often does not have a good correlation with the internal tumor motion, particularly for lung cancer. This study proposes an Advanced Surface Imaging (A-SI) framework to address this issue. In the proposed A-SI framework, a high-speed surface imaging camera consistently captures surface images during radiation delivery, and a CBCT imager captures single-angle X-ray projections at low frequency. The A-SI then utilizes a generative model to generate real-time volumetric images with full anatomy, referred to as Optical Surface-Derived cone beam computed tomography (OSD-CBCT), based on the real-time high-frequent surface images and the low-frequency collected single-angle X-ray projections. The generated OSD-CBCT can provide accurate tumor motion for precise radiation delivery. The A-SI framework uses a patient-specific generative model: physics-integrated consistency-refinement denoising diffusion probabilistic model (PC-DDPM). This model leverages patient-specific anatomical structures and respiratory motion patterns derived from four-dimensional CT (4DCT) during treatment planning. It then employs a geometric transformation module (GTM) to extract volumetric anatomy information from the single-angle X-ray projection. A physics-integrated and cycle-consistency refinement strategy uses this information and the surface images to guide the DDPM, generating high quality OSD-CBCTs throughout the entire radiation delivery. A simulation study with 22 lung cancer patients evaluated the A-SI framework supported by PC-DDPM. The results showed that the framework produced real-time OSD-CBCT with high reconstruction fidelity and precise tumor localization. These results were validated through comprehensive intensity-, structural-, visual-, and clinical-level assessments. This study demonstrates the potential of A-SI to enable real-time tumor tracking with minimal imaging dose, advancing SGRT for motion-associated cancers and interventional procedures.
	
	
	\bigbreak
	\bigbreak

	\noindent 
	\section{INTRODUCTION}
	
	Image-guided radiation therapy (IGRT) integrates imaging modalities during radiotherapy, aiming to ensure that radiation is accurately delivered to the tumor while minimizing exposure to surrounding healthy tissues \cite{IGRT1,IGRT2,IGRT3}. Typically, a four-dimensional computed tomography (4DCT) scan is performed at the radiation treatment planning stage for thoracic and abdominal sites. It reconstructs a 3D CT volume for every respiratory phase to assess respiratory and tumor motion, as well as to aid in target delineation and organs-at-risk (OARs) segmentation for treatment planning of those motion-associated treatment sites. 3D cone beam computed tomography (CBCT), which visualizes the 3D anatomical volume around the tumor region, is an imaging modality commonly used in IGRT for treatment setup purposes. However, as CBCT imaging system has a slow rotation speed of the scanner on the linear accelerator (Linac), the image acquisition process often takes much longer than a respiratory cycle (e.g., approximately one minute for modern Linacs and about 16 seconds for the new HyberSight CBCT imaging system). Consequently, CBCT images are pre-treatment images taken before each radiation delivery fraction.
	
	As a result, CBCT images for the thoracic and abdominal sites are subject to motion artifacts and can only infer the average tumor location to account for the inter-fractional change. The intra-fractional tumor motion is disregarded, such as that caused by breathing. Although the emerging 4D CBCT imaging technology can provide a set of time-resolved 3D CBCT images to assess the tumor motion on the day of treatment, its inferior image quality, prolonged image acquisition time (approximately four minutes), and significantly increased imaging dose hinders its widespread clinical adoption \cite{4DCBCT1}. More importantly, as both CBCT and 4D CBCT are acquired prior to treatment delivery, neither of them can monitor the intra-fractional motion changes during the treatment, which could compromise the accuracy of delivery. 
	
	Although lots of efforts have been devoted to developing high-quality, low-dose, and fast CBCT imaging, including compressed sensing techniques\cite{compress1,compress2,compress4,compress5}, regularization models\cite{regu1,regu2,regu3}, and deep learning models\cite{dl1,dl2,dl3}, these approaches could only reduce the required number of projections by ~60\%. This data sparsity level is far from achieving real-time high-quality volumetric imaging. Moreover, some of these methods require two simultaneous orthogonal kV projections, which are not feasible for most linac systems without an additional imaging system. Recently, patient-specific deep learning models have been proposed for ultra-sparse medical imaging tasks, leveraging patient-specific prior knowledge gained during model training to generate 3D images from single-angle X-ray projections \cite{patredose,patreradio,patreconshen,patstudylei}. For instance, Shen et al.\cite{patreconshen} applied a patient-specific cascade V-shape convolutional neural network (V-net) \cite{Vnet} to map a single 2D projection of either an anterior-posterior view angle or lateral view angle to a corresponding 3D CT image. Lei et al.\cite{patstudylei} developed a novel patient-specific generative adversarial network \cite{ganGoodfellow} integrated with perceptual supervision to derive instantaneous volumetric images from a single 2D projection of any view angle. While Lei et al.\cite{patstudylei} demonstrated the feasibility of using single X-ray projection for real-time in-treatment volumetric imaging and tumor tracking for lung cancer radiotherapy through a simulation study, these methods require constant acquisition of X-ray projections during the treatment, which raise concerns about the imaging dose exposed to patients.  
	
	On the other hand, surface-guided radiotherapy (SGRT) offers a real-time and radiation-free motion monitoring solution to complement current IGRT. It utilizes 2D optical surface images (OSI) to track surface changes during treatment  \cite{SGRTLIU,opticalAl,opticalgie,surfacecarl}. SGRT assumes that respiratory surface topography changes can be a good surrogate for internal tumor motion. By combining the in-treatment surface topography images with the pre-treatment CBCT images as a reference, OSI can enhance IGRT by offering an inferred real-time tumor motion tracking, enabling more accurate radiation delivery. However, this assumption may fail in treatment sites where surface movement may not accurately reflect internal tumor motion \cite{surfaceStanley,surfacewal}. The lack of detailed internal anatomical information in OSI introduces uncertainty in tumor motion tracking for cases with intra-fractional motion changes, particularly in lung, thoracic, and abdominal sites. This limitation restricts SGRT's precision in radiation delivery. Therefore, a new imaging framework that takes advantage of the real-time, radiation-free capabilities of OSI and the detailed anatomical imaging of CBCT could greatly enhance radiation delivery accuracy.
	
	Inspired by the patient-specific single-projection-driven methods \cite{Vnet,Patient4}, we have developed a patient-specific generative adversarial model \cite{ganGoodfellow} to generate 3D CBCT images from optical surface images in our previous study \cite{patresurpan}. In this paper, we aim to further improve the accuracy of the surface-driven 3D image generation by taking advantage of both high-frame and radiation-free surface imaging and the internal anatomy information provided by X-ray projections. Therefore, in this study, we propose an Advanced Surface Imaging (A-SI) framework to guide radiotherapy. The A-SI divides each delivery fraction (one-day delivery) into multiple pre-determined multi-view windows, terms radiation windows, where each window consists of a set of different angular radiation views. It is designed to generate in-treatment real-time cone beam reconstructed volumetric images with entire anatomy from real-time surface topography images, termed Optical Surface-Derived CBCT (OSD-CBCT), for all views across all windows. At the treatment planning stage, surface images, 4DCT, and single-angle cone beam X-ray projections from the 4DCT are collected from each patient. This data is used to train a patient-specific deep learning-based generative model to generate OSD-CBCT from surface images with the assistance of a single-angle cone beam X-ray projection during treatment. In treatment stage, during each radiation window, a high-frequency optical camera continuously captures 2D surface images throughout the entire window. The CBCT imager only takes a single-angle 2D X-ray projection at the starts of every window (the first-view) to minimize imaging dose. For any moment in each delivery window, the generative model then produces OSD-CBCTs in real-time manner from surface images of that moment and the single-angle 2D X-ray projection at the starts of the window. The OSD-CBCTs enable visualization and accurate monitoring of the moving tumor during delivery.
	
	A key technique in this framework is the patient-specific deep learning-based generative model. Due to the lack of internal anatomical information in surface images and the sparse anatomical information in the single-angle cone beam X-ray projection, this model must address the challenge of generating full anatomical images from ultra-sparse information. Motivated by the aforementioned patient-specific projection-driven methods \cite{Patient1,patient2,Patient3, Patient4, Patient5}, we propose a patient-specific model to learn each patient's unique anatomical features and motion patterns for this ultra-sparse generation during the model-training process. To enable the proposed A-SI framework, we introduce a patient-specific physics-integrated consistency-refinement denoising diffusion probabilistic model (PC-DDPM). It leverages the high visual realism and quality of DDPMs \cite{DDPM2,DDPM3,DDPM1} as the generative model. In the A-SI framework, because the surface images are collected throughout the entire delivery, but the projection is only collected at the first-view of each delivery window, the PC-DDPM is specifically designed to generate OSD-CBCT in two scenarios: 1) for each window, the first-view surface topography image has a corresponding single-angle cone-beam X-ray projection; 2) the surface image at subsequent views lacks X-ray projections.  For the first scenario, a physics-integrated (PI) refinement strategy uses two distinct DDPMs: the CBCT-DDPM to generate OSD-CBCT from a single-angle projection and the projection-DDPM to generate full-angle projections. These outputs reciprocally guide each other through a geometric transformation module (GTM) to produce an OSD-CBCT. The GTM also incorporates physics-based geometric knowledge into the generation to enhance the OSD-CBCT's quality. For the second scenario, a cycle-consistency (Cycle-C) refinement strategy uses the first-view OSD-CBCT as a reference to guide the generation of OSD-CBCTs for subsequent views based on a cycle-consistency principle \cite{cyclegan}. In summary, the PC-DDPM can generate OSD-CBCTs at the first view and subsequent views, enabling the A-SI framework to achieve real-time imaging to guide the entire radiation delivery.
	
	The main contributions of this paper are summarized as follows:
	\begin{itemize}
		\item We introduced a novel A-SI framework that generates real-time CBCT (OSD-CBCT) images from surface topography images with the assistance of single-angle cone beam X-ray projections during SGRT. This framework advances SGRT by supporting real-time tumor motion tracking and improves radiation treatment delivery accuracy.
		\item We proposed the PC-DDPM to enable the imaging framework. This model employs PI and Cycle-C strategies to utilize both high-frequency, real-time surface images and low-frequency cone-beam X-ray projections, enhancing the quality of the generated OSD-CBCT images.
		\item We developed a GTM to simulate the cone beam geometric process to guide the PC-DDPM, improving the model's performance.
		\item Our patient-specific approach ensures that the generated OSD-CBCT images are highly relevant and tailored to the specific patient undergoing treatment.
	\end{itemize}
	The rest of this paper is organized as follows: Section 2 reviews related work. Section 3 presents our proposed PC-DDPM model. Section 4 describes the experimental and evaluation setup. Section 5 presents the evaluation results. Section 6 offers a discussion. Finally, Section 7 concludes the paper.
	
	\section{RELATED WORKS}
	\subsection{Ultra-sparse Image Synthesis and Patient-specific Model}
	Recent advancements in ultra-sparse image synthesis have focused on reconstructing 3D images from a few 2D images. This approach is particularly relevant in object reconstruction \cite{rec1,rec2,rec3,rec4,twoprojection1,twoprojection2,twoprojection3} and volume rendering \cite{render1,render2,render3}. In the 3D medical image domain, ultra-sparse synthesis is mainly achieved through patient-specific models since medical images require extremely high accuracy for clinical purposes. Patient-specific principle can be adapted to different types of neural networks. Existing works include patient-specific cascade V-shape convolutional neural networks (V-net) \cite{Patient1,patient2,Patient5} and patient-specific generative adversarial networks (GANs) \cite{patient3,Patient4}. These models reconstruct 3D volumes from one or two 2D X-ray projections. Collectively, these research endeavors affirm the potential of using patient-specific models to solve ultra-sparse synthesis challenges and enable the A-SI framework. They also lay the groundwork for integrating a more sophisticated generative model than V-net and GANs using patient-specific principles to achieve accurate OSD-CBCT generation.

	\subsection{Denoising Diffusion Probabilistic Models}
	DDPMs have advanced in medical image-to-image translation. Recent research has focused on their potential across various imaging modalities \cite{medicalddpm1,medicalddpm3,medicalddpm4,medicalddpm10}. Efforts also target enhancing DDPM efficiency \cite{medicalddpm5,medicalddpm6} and refining learning processes for superior image accuracy \cite{medicalddpm8,ddpmrepaint,medicalddpm9,medicalddpm11}. Despite these advancements, existing DDPMs are unsuitable for our framework. Firstly, our DDPM must accept both surface images and inconsistently collected single-view cone-beam projections. It needs to adapt to two scenarios: with both surface images and projections, or with only surface images. Current DDPMs lack this adaptability. More specifically, in the first scenario, the generated OSD-CBCT must not only exhibit excellent visual quality but also adhere to real-world cone beam geometric transformations during the projection-to-CBCT reconstruction process. Current DDPMs lack mechanisms to ensure synthesis obeys these physical transformations. In the second scenario, the surface image alone may not provide any internal anatomical information. A mechanism is needed to utilize the single-view projection, although in a different view, to convey the anatomical information to the current view OSD-CBCT. Existing DDPMs do not have such mechanisms. These challenges motivate the development of the PC-DDPM, designed to enable the proposed A-SI framework.

	\noindent 
	\section{Methodology}
	
	We proposed A-SI framework, as shown in Fig. 1, utilizing PC-DDPM from an input Gaussian noise to generate OSD-CBCT as output. OSD-CBCT generation iteratively removes noise over multiple timesteps, transforming an initial pure noise volume into high-quality OSD-CBCT in Fig. 1A. In each radiation window, we only collect a single-view projection at the beginning. Accordingly, the PC-DDPM is designed to generate OSD-CBCT generation under two imaging scenarios. The initial is the “first-view” scenario is there existed a single-angle projection (X-ray). The second is the “subsequent view” scenario when we stop collecting projections. In each generation step, depending on the A-SI framework's imaging scenario, we apply PI refinement (Fig. 1B) to generate first-view OSD-CBCT with projection and Cycle-C refinement (Fig. 1C) for generating subsequent-view OSD-CBCT without projection. 
	
	\begin{figure}
		\centering
		\noindent \includegraphics*[width=8in, height=6in, keepaspectratio=true]{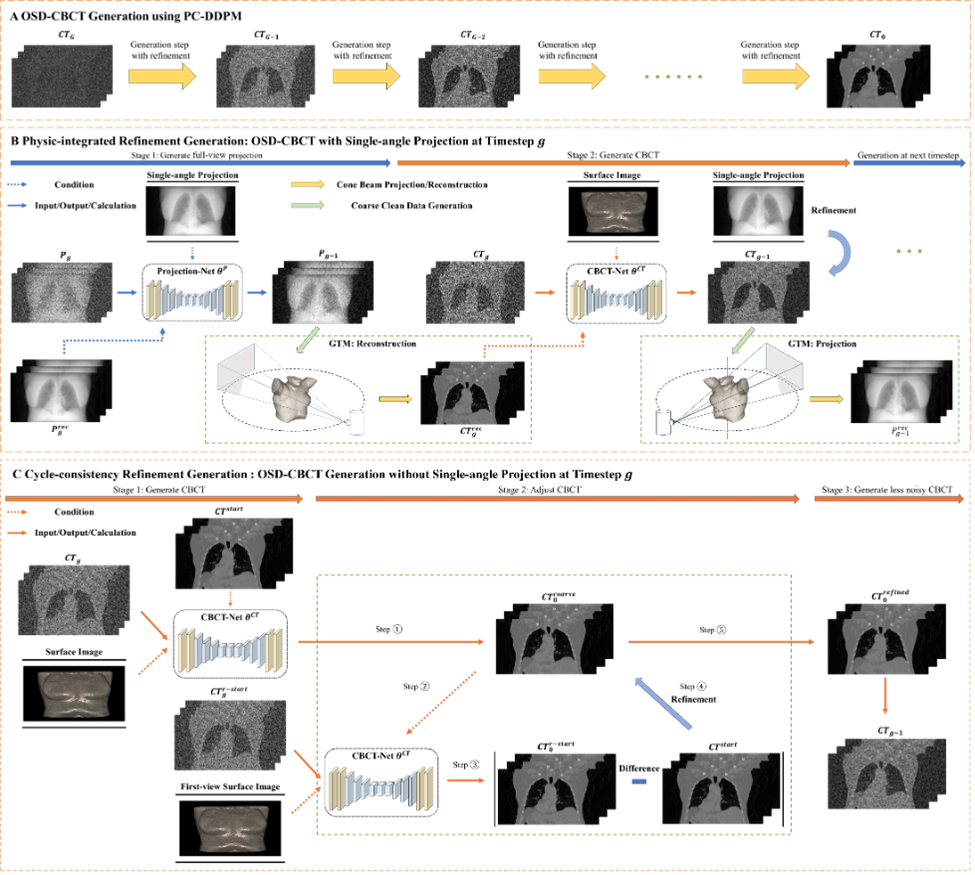}
		
		\noindent Figure 1: (a) Overview of the diffusion process: iteratively generating OSD-CBCT from Gaussian noise using different refinements for different OSD-CBCT. PI refinement is applied for the first-view OSD-CBCT, and Cycle-C refinement is used for subsequent views. (b) PI refinement: Surface images and single-angle projections generate full-view projections, which are denoised, geometrically transformed (via FDK algorithm), and reconstructed into CBCT. This CBCT is used to enhance Projection-Net for the next timestep. (c) Cycle-C refinement: Denoises subsequent-view OSD-CBCT using surface images and first-view CBCT as anatomical reference. A cycle-consistency step refines the OSD-CBCT by comparing regenerated and original first-view CBCT, improving the current view.
	\end{figure}
	
	The PI refinement for the first-view scenario, where at timestep g, the projection-DDPM generates less noisy full-angle projections conditioned on the single-angle projection, is depicted in Fig. 1B. These projections after projection-DDPM feed into the GTM to reconstruct CBCT, levitate the CBCT-DDPM to producing a refined OSD-CBCT with less noise. Then, the refined OSD-CBCT passes through the GTM to generate full-angle projections, enhancing the projection-DDPM for the next timestep in this cycle-generation manner. In this way, the first-view OSD-CBCT is directly guided by the corresponding projection.
	
	The Cycle-C refinement for the subsequent view scenario, which uses only the surface image to generate high-quality OSD-CBCTs at each timestep, is illustrated in Fig. 1C. Initially, the first-view OSD-CBCT with projection enhances the CBCT-DDPM, generating a coarse current subsequent-view OSD-CBCT. This current OSD-CBCT then regenerates the first-view OSD-CBCT via the same network. The difference between the regenerated and the original first-view OSD-CBCT refines the current OSD-CBCT. In this way, the current OSD-CBCT can be indirectly guided by the first-view projection.

	\subsection{Diffusion Model Formulation}
	In this work, we utilize a conditional DDPM (C-DDPM) for our PC-DDPM. C-DDPMs generate data through a forward diffusion and a reverse denoising process. In forward diffusion, data points, denoted as $X_0$ (OSD-CBCT volume for CBCT-DDPM or full-angle projection for Projection-DDPM), are corrupted by adding random noise over N timesteps, becoming pure Gaussian noise. Conversely, the reverse denoising process starts with pure noise and iteratively removes it, mimicking reverse diffusion process. In our modified reverse process, we are enhancing the standard reverse process by incorporate additional conditions, denoted as y (i.e., surface image for CBCT-DDPM or single-angle projection for Projection-DDPM), to guide denoising. This ensures that generated data adheres to the conditions provided. The reverse process is called denoising for training data and generation when used to create new OSD-CBCT.
	
	Forward diffusion gradually adds noise to initial data $X_0$ by a stochastic differential equation (SDE) \cite{ddpmsong}:
	\begin{equation}
		dX=-\frac{1}{2}\beta(t)X\cdot+\sqrt{\beta(t)}\cdot\omega
	\end{equation}
	where $\beta(t)$ is the pre-determined noise schedule, which controls how much noise is added at each timestep, namely the variance, while N(0,I) denotes Gaussian noise. The process is discretized over N timesteps, yielding $X_n$ of noisy data:
	\begin{equation}
		X_{n} = \sqrt{1 - \beta(n)} X_{n-1} + \sqrt{\beta(n)} \epsilon_n
	\end{equation}
	with $\epsilon_n \sim \mathcal{N}(0, I)$ is the mean of random noise added at each timesteps. To generate $X_{n+1}$ directly from $X_0$where $\alpha_{n} = \prod_{s=1}^{n} (1 - \beta_s)$, we have:
	\begin{equation}
		X_{n} = \sqrt{\alpha_{n}} X_0 + \sqrt{1 - \alpha_{n}} \epsilon_n
	\end{equation}
	Then the reverse process removes noise to recover the original data by the C-DDPM:
	
	\begin{equation}
		dX_n = \left[-\frac{1}{2} \beta(n)X -\beta(t) \nabla_{X_t} \log p(X_t \mid y)\right] dg + \sqrt{\Sigma} d\tilde{W}_n
	\end{equation}
	where $\nabla_{X_t} \log p(X \mid y)$ is the score function representing the gradient of the logarithm probability distribution of the noisy data $X_t$ given the condition y for the C-DDPM. $\Sigma$ is a pre-determined variance, and $d\tilde{W}_t$ is the Wiener process that model the random moves \cite{ddpmsong}. Exact score function computation is intractable, so we train a neural network to estimate coarse clean data $X_0^{coarse}$, taking inputs of $X_n$ using a mean absolute error conditioning on $y$. And the $L(X_0^{coarse},n,y)$ represent the loss function to estimate the $X_0^{coarse}$  from noisy data at timestep n:
	\begin{equation}
		L(X_0^{coarse},n,y) = \mathbb{E}_{X_0, n} \left[ \left\| X_0 - X_0^{coarse}(X_n, n, y) \right\|_2 \right]
	\end{equation}
	where $\mathbb{E}_{X_0, n}$ denotes the expectation over all timesteps n (the joint distribution of the clean data$X_0$ and the noise level at timestep n). With $X_0^{coarse}$, the score function is:
	\begin{equation}
		\nabla_{X_t} \log p(X_t \mid y)\approx s(X_n, x^{coarse}_0 n, y) = -\frac{X_n - \sqrt{\alpha_n} X_0^{coarse}}{\sqrt{1 - \alpha_n}}
	\end{equation}
	The score function is a defining feature of our method, crucial for guiding the refinement of both CT and projection data, will also play a significant role in the generation process, where it helps reduce noise and improve overall accuracy. This will be explored in detail in the generation part of our paper. In classic diffusion model, the variance $\Sigma$ is defined as $\beta_n^2$ requiring the number of timesteps for generation to match those in training. Instead, we estimated the variance using an interpolation coefficient $v(X_n,n,y)$ to balance the $\beta_n$  and estimate variance \cite{ddpmnichol}:
	
	\begin{equation}
		\begin{aligned}
			\Sigma(X_n, n, y) &= \exp (v(X_n, n, y) \log(\beta_n) \\
			&+ (1 - v(X_n, n, y)) \log(\frac{1 - \prod_{i=1}^{n-1} \alpha_i}{1 - \prod_{i=1}^{n} \alpha_i} \beta_n))
		\end{aligned}
	\end{equation}
	Also using this approach, the reverse diffusion process requires significantly fewer timesteps than in training, enhancing generation speed. Variance optimization involves minimizing the Kullback-Leibler (KL) divergence between the ground truth distribution $\nabla_{X_t} \log p(X_t \mid y)$ with variance  $\beta_n^2$  and the estimate score function $s(X_n,n,y)$ and variance $\Sigma(X_n,n,y)$ across all $X_0$ and timesteps n with the condition y::
	\begin{equation}
		\begin{aligned}
			L(\Sigma,n,y) &= \mathbb{E}_{X_0, n} [\text{KL}(N\left(-\frac{1}{2}\beta_n X - \beta_n \nabla_{X_n} \log p(X_n \mid y), \beta_n^2\right) \\
			&\parallel N(-\frac{1}{2}\beta_n X - \beta_n s(X_n, n, y), \Sigma(X_n, n, y)))]
		\end{aligned}
	\end{equation}
	The final loss function for C-DDPM combines two terms both under condition y: loss for estimating the coarse clean data $X_0^{coarse}$  from noisy data $X_n$ and variance loss from Eqn. 8. And the $\lambda_1$ is a constant empirically selected as 0.05 that balances the two loss terms:
	\begin{equation}
		L_{DDPM}(s,\Sigma) = L(X_0^{coarse},n,y) + \lambda_1 L(\Sigma,n,y)
	\end{equation}
	The final generation reverse diffusion process utilizes a discrete-time version of Eqn. 4 (which is DDPM) to calculate iteratively removes noise from $X_n$ to $X_(n-1)$ at timestep n-1::
	\begin{equation}
		X_{n-1} = \frac{1}{\sqrt{1-\beta_n}}(X_n+\beta_ns(X_n,n,y)) + \sqrt{\Sigma(X_n, n, y)} \epsilon_n
	\end{equation}
	where $\epsilon\sim N(0,I)$.Using Eqn. 10, CBCT-DDPM can generate OSD-CBCT from pure Gaussian noise ($X_N$) and iteratively refining it with the score function and variance estimation until it produces high-quality clean images. For CBCT-DDPM, we define the condition is $y_{CT} = [S \mid CT_{s-rec}]$, where $S$ indicates the surface volume converted from the optical surface mesh, while $\mid$ indicates concatenation. $CT_{s-rec}$ to a CBCT reconstructed from the single-angle projection using GTM. Similarly, Projection-DDPM can generate full-angle projections from pure Gaussian noise ($X_n$), starting with a single-angle projection as the condition and refining it using the reverse diffusion process. For Projection-DDPM, the condition is $y_{P} = [P_{s-reshape} \mid I]$, where $P_{s-reshape}$ is the single-angle projection duplicated 360 times in a new depth dimension to match the targeted full-angle projections, and I is an identity matrix.

	\subsection{Physics-integrated Refinement}
	In the previous section, we outlined the training process for CBCT-DDPM to generate CBCT images and Projection-DDPM to synthesize full-angle projections. For the first-view CBCT, where both surface images and projections are available, the projection provides internal anatomical details and ensures adherence to physical geometry. To further enhance the accuracy of CBCT generation, we employ dual-domain DDPMs, leveraging both the CBCT and full-angle projection data to guide the process based on the first-view projections.
	
	Each timestep has two stages as shown in Fig. 1B. In stage 1, Projection-Net refines a coarse clean full-angle projection by utilizing a single-angle projection and the projection reconstructed by OSD-CBCT from the previous timestep as conditioning inputs. In stage 2, the GTM (details are shown in Section 3.4), a deep-learning framework using cone-beam geometry, reconstructs the coarse projection into a CBCT.  CBCT-Net generates a less noisy OSD-CBCT by utilizing the condition of single-angle projection and surface images as inputs, while also incorporating the reconstructed CBCT from the GTM for additional guidance for further enhanced accuracy. The OSD-CBCT is further refined by the corresponding single-angle projection. The refined OSD-CBCT then assists Projection-Net in the next stage. These DDPMs enhance each other in a cycle-generation process with the GTM.
	
	We define the following term: the Projection-Net as $\theta^{P}$ and CBCT-Net as $\theta^{CT}$. Then, we denote the true full-angle projections as $P_0$, , the estimated less noisy projections as $P_{n-1}$, and and the coarse clean projections as $P^{coarse}_0$. The CBCT reconstructed from the coarse clean projections is denoted as $CT^n_{rec}$. TWe set the true OSD-CBCT for $CT_0$, the estimated less noisy OSD-CBCT for $CT_{n-1}$, and the coarse clean OSD-CBCT for $CT^{coarse}_0$. The full-view projection projected from the coarse clean CBCT is $P^{rec}_n$. These definitions will be used to explain the training and generation formulations.
	
	\subsubsection{Training}
	Starting with the Projection-DDPM, the Projection-Net is trained to estimate $P^{coarse}_0$ optimized by loss function Eqn. 5. Note that Projection-Net outputs coarse full-angle projection using a single-angle projection and the previous OSD-CBCT reconstruction as inputs. Then we employ GTM to reconstruct $P^{coarse}_0$ into $CT^{rec}_n$. This reconstructed CBCT should minimize the difference with $CT_0$ if $P^{coarse}_0$ adheres to the cone beam physics geometry. For this purpose, $CT^{rec}_n$ is optimized by a CBCT-reconstruction loss:
	\begin{equation}
		L_{rec}(CT^{rec}_n) = \mathbb{E}_{CT_0, n} \left[ \| CT_0 - GTN(P_0^{coarse}) \|_2 \right]
	\end{equation}
	Additionally, $CT^{rec}_n$ serves as an information-rich anatomical condition, enabling the CBCT-DDPM to generate high-quality OSD-CBCT. By inputting a new condition $y_{rec} = [S \mid CT^{rec}_n]$into the CBCT-Net, we generate more accurate and less noisy OSD-CBCT. The generation's score function $s_{CT-rec}$ and variance $\Sigma_{CT-rec}$ with this new condition are optimized by $L(X_0^{coarse},n,y_{CT-rec})$ and $L(\Sigma_{CT-rec},n,y_{CT-rec})$ similar to Eqn. 5 and 8, and the overall loss function is:
	\begin{equation}
		L_{CT-rec}(s_CT,\Sigma_{CT-rec}) = L(CT^{coarse},n,y_{CT-rec})+\lambda_1L(\Sigma_{CT-rec},n,y_{CT-rec})
	\end{equation}
	The final loss function for CBCT generation combines the optimization using both conditions $y_CT$ and $y_{CT-rec}$, where CBCT-Net can handle both conditions. Further combining with the Eqn. 11, the CBCT-Net and Projection-Net are jointly optimized by::
	\begin{equation}
		\begin{aligned}
			L_{CT}=\gamma_1L_{DDPM}(s_{CT},\Sigma_{CT})+\gamma_2L_{CT-rec}(s_{CT-rec},\Sigma_{CT-rec})+\gamma_3L_{rec}(P_0^{coarse})
		\end{aligned}
	\end{equation}
	Where weighting constants $\gamma_1$, $\gamma_2$and $\gamma_3$ are empirically chosen as 0.45, 0.45, and 0.1, respectively. 
	
	Similarly, CBCT-Net reduces noise in OSD-CBCT generation by using single-angle projection and surface images as primary input. Then it is trained to estimate $CT_0^{coarse}$ , which can be converted into full-angle projections using the GTM that provide additional guidance for enhanced accuracy. This conversion supervises the Projection-DDPM to calculate the full-angle projections’ score function $s_{P-rec}$ and variance $\Sigma_{P-rec}$, using a new condition $y_{P-rec}= [P_{s-reshape} \mid P_n^{rec}]$, which is optimized as:
	\begin{equation}
		\begin{aligned}
			L_{P} &= \gamma_1(L_{DDPM}(s_p,\Sigma_p))  + \gamma_2L_{P-rec}(s_{P-rec},\Sigma_{P-rec}) \\
			&+ \gamma_3(L_{rec}(CT^{coarse}_0)
		\end{aligned}
	\end{equation}
	This reciprocal supervision between these two DDPMs during training phases fosters mutual improvements for each DDPM. The final loss function for the PI refinement is:
	\begin{equation}
		L_{PI} = 0.5(L_{CT} + L_{P})
	\end{equation}
	The number of training timesteps $N$ is selected as 1000.
	
	\subsubsection{Generation}
	During the generation process, G generation timesteps are proposed to iteratively use the CBCT- and Projection-DDPM, generating a final OSD-CBCT image. These timesteps do not necessarily match the number of training timesteps  \cite{ddpmnichol}. In our method, we evenly space the training steps between 1 and N by G timesteps. This allows us to obtain the corresponding noise schedule $\beta_g$ to perform the reverse generation.
	
	Beginning from timestep $G$, $CT_{g-1}$ and $P_{g-1}$ are sampled from $CT_g$ and $P_g$ independently, without using $CT^{coarse}_0$, $P^{coarse}_0$, and GTM. At these stages, they are still noisy and provide limited information. Starting From timestep $G-l$, we initiate the PI refinement strategy in two stages, as illustrated in Stage 1 of Fig. 1B. In stage 1,  $P_0^{coarse}$  is directly generated by the Projection-Net, supported by the $CT_g$ (the reconstructed CBCT obtaining from the previous timestep by solely using CBCT-DDPM):
	\begin{equation}
		P^{coarse}_0 = \theta^{P}(P_g, g, y_{rec})
	\end{equation}
	We also generate the estimated less noisy projections $P_{g-1}$, where the score functions $s(\cdot)$ is calculated using Eqn. 6.:
	\begin{equation}
		P_{g-1} =\frac{1}{\sqrt{1-\beta_g}}(P_g+\beta_gs(P_g, P_0^{coarse}, y_{P-rec})) + \sqrt{\Sigma(X_g, g, y_{P-rec})} \epsilon
	\end{equation}
	
	In stage 2, $P^{coarse}_0$ is reconstructed via GTM into  $CT^g_{rec}$ , forming the condition  $y_{CT-rec}=[S\mid CT_g^{rec}]$ to further enhance the CBCT-DDPM at the same timestep::
	\begin{equation}
		CT^{coarse}_0 = \theta^{CT}_0(CT_g, g, y_{CT-rec})
	\end{equation}
	Additionally, we aim to enhance the OSD-CBCT image generation by incorporating true single-angle projection as a data consistency constraint. This ensures that the generated OSD-CBCT aligns with both the projection data and the physical geometry between the projection and the OSD-CBCT. Following Chung et al.'s work \cite{ddpmchung}, the single-view projection $P^s$ enforces the $CT^{coarse}_0$ to follow cone-beam imaging physical geometry, where $\nabla$ denotes the gradient through the GTM:
	\begin{equation}
		\begin{aligned}
			CT^{refined}_0 &= CT^{coarse}_0 \\
			&- \frac{1}{\Sigma_{CT-rec}}\nabla_{CT^{coarse}_0} \| P^s - GTM(CT^{coarse}_0) \|_2
		\end{aligned}
	\end{equation}
	where $\nabla$ denotes the gradient.
	Eventually, the refined $CT^{refined}_0$ is used to calculate a score function $s_{refined}$ using Eqn. 6, and then $CT_{g-1}$:
	\begin{equation}
		CT_{g-1} = \frac{1}{\sqrt{1-\beta_g}}(CT_g + \beta_g s_{refined}(CT_g,CT^{refined}_0,g,y_{CT-rec})) + \sqrt{\Sigma(CT_g, g, y_{CT-rec})} \epsilon
	\end{equation}
	The refined $CT^{refined}_0$ is also utilized in the Projection-DDPM's generation at the next timestep. By recursing this cycle generation, we eventually obtain  $CT_0$ as the final CBCT. Empirically, $g$ is selected as 10, and $l$ is selected as 5.

	\subsection{Cycle-consistency Refinement}
	In subsequent views, where only the surface image is available, projections are missing, limiting the available anatomical information. Accordingly, the Projection-DDPM and GTM are disable in this phase. However, the previously OSD-CBCT generated from the first view can serve as a reliable reference for internal anatomy, as it includes X-ray projection data. In this perspective, the subsequent-view OSD-CBCT are indirectly guided by the first-view projection. To enable the first-view OSD-CBCT to enhance the current phase OSD-CBCT, we use the cycle-consistency principle  \cite{cyclegan}: if data A can generate B, then a correct B should be able to regenerate an A identical to the original A. In our work, A is the first-view OSD-CBCT with projection, and B is the current-view OSD-CBCT. The full Cycle-C refinement in each generation step can be summarized in three stages.
	
	In stage 1, the first-view OSD-CBCT supports the CBCT-Net in generating a coarse clean current OSD-CBCT from the current surface image and a noisy OSD-CBCT, as shown in Fig. 1C. In stage 2, the coarse OSD-CBCT is refined using the cycle-consistency principle, and the detailed steps involved will be discussed in the following. We start at a current coarse OSD-CBCT generated from stage 1. Next, the coarse OSD-CBCT is input into the same CBCT-Net. Then, CBCT-Net takes the first-view surface image and the noisy CBCT as inputs, and with the support of the coarse OSD-CBCT, generates a clean first-view OSD-CBCT. After that, we calculate the differences between the regenerated and given first-view OSD-CBCT. These differences refine the current-view OSD-CBCT, ensuring it can accurately regenerate a first-view OSD-CBCT that is identical to the original first-view OSD-CBCT. This cycle-consistency principle \cite{gankarras} allows us to generate a refined current clean OSD-CBCT. In stage 3, the current refined clean OSD-CBCT is used to calculate the final less noisy OSD-CBCT.
	
	For the generation, we term  \(CT^{start}\) is the first-view OSD-CBCT with projection;  \(CT_n\) denotes the current-view noisy OSD-CBCT at timestep  $n$. \(CT^{coarse}_0\) represents the clean current-view OSD-CBCT generated in stage 1;  \(CT^{r-start}_n\) is the noisy first-view OSD-CBCT, and \(CT^{r-start}_0\) is the clean first-view OSD-CBCT generated from the current-view OSD-CBCT; the refined clean OSD-CBCT is denoted as \(CT^{refined}_0\). Finally, \(CT_{n-1}\) is the less noisy current OSD-CBCT calculated from \(CT^{refined}_0\).
	
	\subsubsection{Training}
	During training, we first use \(CT^{start}\), which already generated in the first view of the radiation window, as a condition to construct \(y_{start} = [S \mid CT^{start}]\). This trains the CBCT-Net to obtain the current-view  \(CT^{coarse}_0\). Similar to Eqn. 9, the CBCT-Net is optimized by additional loss functions to learn the score function $s_{start}$ and variance $\Sigma_{start}$ with the new condition:
	\begin{equation}
		\begin{aligned}
			L_{start}(s_{start},\Sigma_{start}) = L(CT_0^{coarse},n,y_{start})+\lambda_1L(\Sigma_{CT},n,y_{start})
		\end{aligned}
	\end{equation}
	We then duplicate a frozen CBCT-Net to regenerate a first-view OSD-CBCT. We firstly obtain a noisy first-view CBCT $CT_n^{r-start}$ via Eqn. 3. Then we set another new condition $y_{cycle} = [S\mid CT_0^{coarse} ]$, which trains the model to regenerate a first-view $CT_0^{r-start}$ from $CT_n^{r-start}$. The CBCT-Net is optimized to force the regenerated OSD-CBCT to be identical to the original OSD-CBCT using a cycle-consistency loss function:
	\begin{equation}
		\begin{aligned}
			L_{cycle}(CT^{coarse}_0) = \mathbb{E}_{C^{r-start}, n} \left[ \left\| CT_0^{r-start}(CT_n^{r-start}, n, y_{cycle})-CT^{start} \right\|_2 \right]
		\end{aligned}
	\end{equation}
	The loss function for the Cycle-C refinements is:
	\begin{equation}
		\begin{aligned}
			L_{cycle-C} = L_{start}(s_{start},\Sigma_{start})+\lambda_2L_{cycle(CT_0^{coarse})}
		\end{aligned}
	\end{equation}
	Where weighting constants $\lambda_2$ is empirically selected as 0.1. The final loss function for the CBCT-Net, including both the PI and Cycle-C refinements, is:
	
	\begin{equation}
		\begin{aligned}
			L_{total} & = 0.5(L_{CT} + L_{Cycle-C})
		\end{aligned}
	\end{equation}
	
	\subsubsection{Generation}
	During the generation phase, as shown in Fig. 1C, we begin by generating the current-view OSD-CBCT without projection with the condition  \(y_{start}\) using the generation timestep $g$ as step 1:
	\begin{equation}
		\begin{aligned}
			CT^{coarse}_0 = \theta^{CT}(CT_g, g, y_{start})
		\end{aligned}
	\end{equation}
	Then, \(CT^{coarse}_0\) is used to construct $y_{cycle}  =[S\mid CT_0^{coarse}]$  to regenerate the first-view OSD-CBCT as step 2:
	\begin{equation}
		\begin{aligned}
			CT^{r-start}_0 = \theta^{CT}(CT^{r-start}_g, g, y_{cycle})
		\end{aligned}
	\end{equation}
	Following this, $CT_0^{coarse}$  is further refined to minimize the discrepancy between $CT^{start}$  and $CT^{r-start}$  as step 3 to 5 in Fig. 1C, following the data-consistency principle from Eqn. 20:
	\begin{equation}
		\begin{aligned}
			CT^{refined}_0 &= CT^{coarse}_0 \\
			& - \nabla_{CT^{coarse}_0} \| CT^{r-start}_0 - CT^{start} \|_2
		\end{aligned}
	\end{equation}
	Eventually, the refined \(CT^{refined}_0\) is used to calculate a score function \(s_{refined}\) using Eqn. 6, and then to generate \(CT_{g-1}\):
	\begin{equation}
		\begin{aligned}
			CT_{g-1} = \frac{1}{\sqrt{1-\beta_g}}(CT_g + \beta_g s_{refined}(CT_g,CT_0^{refined},g,y_{start})) + \sqrt{\Sigma(CT_g, g, y^{start})} \epsilon
		\end{aligned}
	\end{equation}
	The generation is repeated until $g$ reaches 0 to obtain the final OSD-CBCT $CT_0$.
	
	\subsection{Geometric Transformation Module (GTM)}

	The GRM simulates a CBCT imaging system, comprising projection and reconstruction phases. In the projection phase, a 3D volume is forward-projected into 2D projections at different projection angles. In the reconstruction phase, these 2D projections are back-projected and reconstructed as a 3D volume. For this study, we employed the FDK algorithm for 3D reconstruction and a standard CBCT projection. Projections are taken from 1 to 360 degrees in the axial plane. The transformation parameters follow the physical model of our X-ray imaging scanner. This system features a detector panel with \(768 \times 1024\) pixels (0.78 x 0.78 $mm^2$ pixel spacing), a 1500 mm source-to-isocenter distance, and a 1000 mm source-to-object distance. It is crucial to note that both forward projection and backward projection are differential and frozen operations. This allows them to be integrated with network modules, enabling end-to-end optimization. This approach maintains fidelity to the physical imaging model while ensuring seamless integration with the computational framework.
	
	\subsection{Network Architecture}
	The network in PC-DDPM is encoder-decoder architecture follow Pan et al.’s work \cite{medicalddpm10}. It reduces spatial information while increasing latent feature representation, then restores the spatial domain. The input first passes through convolutional layers ($3 \times 3 \times 3$ and $1 \times 1 \times 1$) to learn early features, which are downsampled by a factor of 2 in each block. The encoder consists of two downsampled convolutional blocks that capture early local characteristics from high-resolution inputs, followed by seven sequential downsampled Swin-transformer blocks to learn global information from low-resolution features. Three middle Swin-transformer blocks, without downsampling, compute global characteristics.
	The decoder consists of seven upsampled Swin-transformer blocks and two final upsampled convolutional blocks to restore features to their original resolution. The final features pass through two convolutional layers: one estimates noise, the other estimates the variance interpolation coefficient. Timestep encoding uses sinusoidal embedding (SE)  \cite{ddpmho}, with a maximum period of \(10^6\) and a feature dimension of 128. These embeddings are used in all blocks for further computations. 
	
	Residual connections \cite{res} enhance network stability, while shortcut connections  \cite{unet} link each encoder block to its corresponding decoder block at the same resolution level, facilitating high-resolution information transfer and improving estimation accuracy.
	
	\section{Experiments}
	All experiments were implemented using the PyTorch framework in Python 3.8.11 on a workstation running a Linux system with a single NVIDIA A100 GPU with 80GB memory. The CBCT-Net and Projection-Net were jointly trained for 650 epochs with a batch size of 2 using the AdamW optimizer at a learning rate of $2\times10^{-5}$.
	
	\subsection{Datasets}
	We conducted a retrospective study with a cohort of 22 lung cancer patients from our institutional database. These patients underwent radiotherapy and had 4DCT scans during their CT simulation. The median age of the patients was 77 years, with an age range of 54 to 98 years. Patients received stereotactic body radiation therapy (SBRT) with varying prescription doses: 30 Gy in 3 fractions (n=2), 34 Gy in 5 fractions (n=1), 40 Gy in 5 fractions (n=1), 48 Gy in 4 fractions (n=20), 50 Gy in 5 fractions (n=15), 54 Gy in 3 fractions (n=9), and 60 Gy in 8 fractions (n=2). Each 4DCT image set consisted of 10 sets of 3D CT images, corresponding to the 0\% to 90\% respiratory phases.  Images were acquired from a Siemens SOMATOM Definition AS scanner at 120 kV using the Siemens Lung 4D CT protocol (Syngo CT VA48A) with a pitch of 0.8 and a Bf37 reconstruction kernel. The image resolution was $512 \times 512 \times (133-168)$ with a voxel size of $0.9756 \times 0.9756 \times 3 mm^3$.
	
	Independent experiments were conducted for each patient. In each experiment, eight respiratory phases were used for training, one for validation, and the remaining phase for inference. The 2D CBCT projection data were simulated by projecting each 3D phase CT image from 360 angles, using the geometry of the onboard cone beam X-ray system. Each CT included manual segmentation of the gross tumor volume (GTV) and vital structures (esophagus, heart, lung, and spinal cord), contoured by board-certified radiation oncologists. The CBCT images were resampled to achieve a resolution of \(1 \times 1 \times 3 \ \mathrm{mm}^{3}\). All volumes were centered and boundary-cropped, standardizing them to a uniform resolution of \(512 \times 512 \times 128\). For both training and inference, voxel intensities were normalized within the range of \([-1, 1]\), while the intensity levels of the OSD-CBCT were converted back to Hounsfield Units (HU). Similarly, the intensities of the projections were normalized to the \([-1, 1]\) range, with the synthetic projections being restored to their original intensity levels.
	
	\section{Evaluations and Results}
	The visual comparison results and difference maps between the ground truth and the generated first- (with projection, so directly guided by first-view X-ray projection) and subsequent-view (without projection, so indirectly guided by projection) OSD-CBCT, set within [-1024,1650] HU and [0,800] HU windows, are presented in Fig. 2. The OSD-CBCT generated from PC-DDPM, under both imaging scenarios, achieves a realistic visual appearance with accurate anatomical details and minimal differences from the ground truth OSD-CBCT. Importantly, within the red box, the tumor in the generated OSD-CBCT is realistically represented with correct anatomical shape and accurate location. This indicates that the generated OSD-CBCT can accurately demonstrate tumor motion, supporting the proposed A-SI framework.
	
	\begin{figure}
		\centering
		\noindent \includegraphics*[width=6.50in, height=4.20in, keepaspectratio=true]{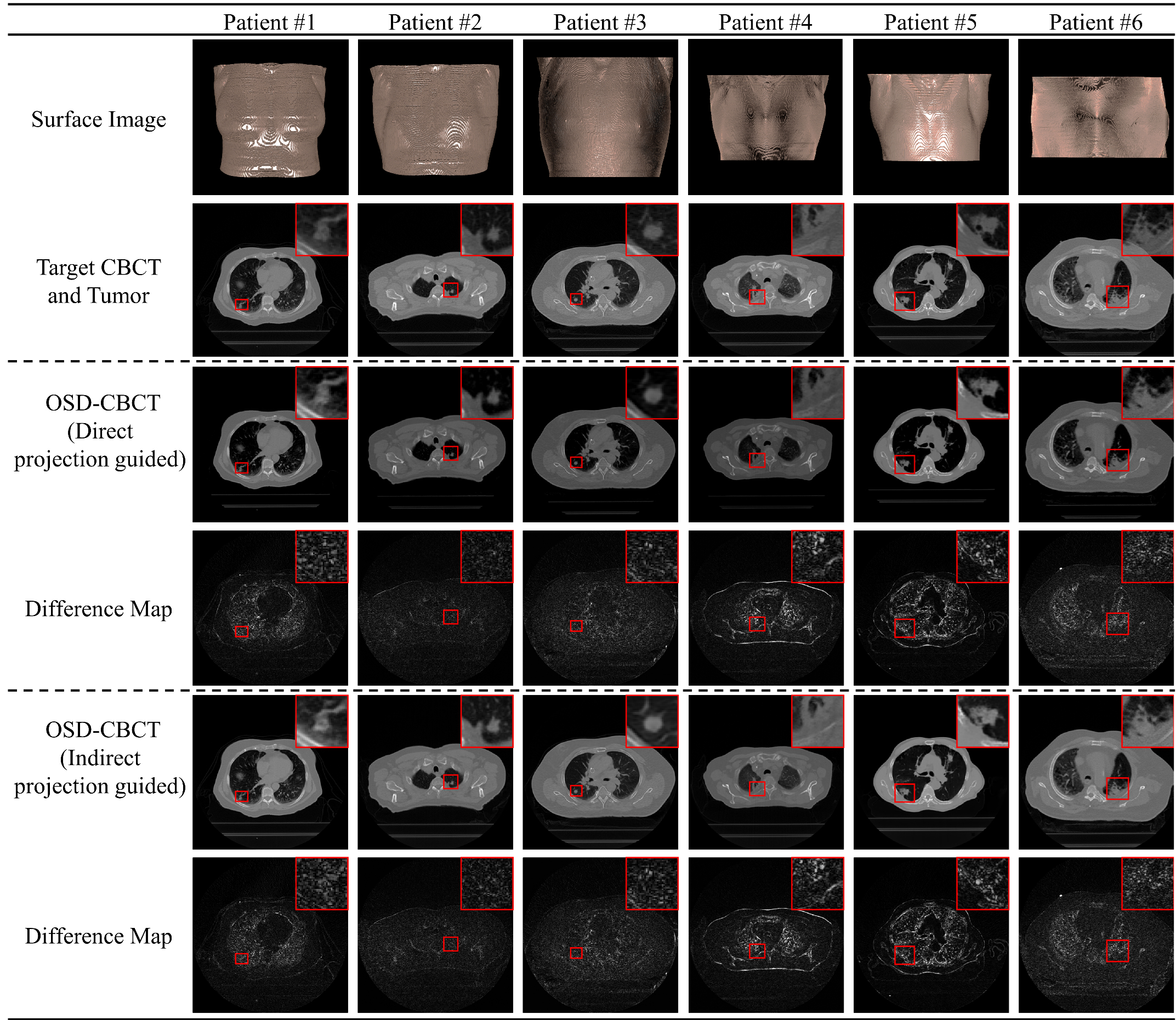}
		
		\noindent Figure 2: Six visual examples of first- (directly guided by first-view X-ray projection) and subsequent-view (indirectly guided by first-view X-ray projection) OSD-CBCT. The first row shows surface images captured by the optical camera. The second row displays the target OSD-CBCT with single-angle projection with zoomed-in tumor regions in the red boxes. The third row presents generated first-view OSD-CBCTs and zoomed tumor regions in each radiation window of the Advanced Surface Imaging (A-SI) framework. The fourth row shows the difference maps between target and generated first-view OSD-CBCTs. The fifth row presents subsequent-view generated OSD-CBCTs without single-angle projection with zoomed tumor regions, and the sixth row shows the corresponding difference maps.
	\end{figure}
	
	\subsection{Evaluation 1: OSD-CBCT Quality Analysis and Metrics}
	To assess the quality of the OSD-CBCT, we used the 4DCT to train the PC-DDPM and simulate ground truth OSD-CBCT to evaluate the first-view and non-first-view (subsequent-view) OSD-CBCT from both imaging scenarios, with and without a single-angle X-ray projection. The first-view OSD-CBCTs were generated independently using X-ray projections acquired at \(0^\circ\), \(30^\circ\), \(60^\circ\), ...,\(360^\circ\) (every \(30^\circ\)). The performance was then evaluated by calculating the average performance across all OSD-CBCT datasets. For the subsequent-view OSD-CBCT, the validation phase served as the reference in the Cycle-C refinement strategy, to generate the inference OSD-CBCT. We define the first-view OSD-CBCT as direct projection guided OSD-CBCT, and the subsequent-view as indirect projection guided OSD-CBCT. The evaluation uses six distinct metrics. These metrics determine the global accuracy of the entire OSD-CBCT. The mean absolute error (MAE) and peak signal-to-noise ratio (PSNR) evaluate the quality of the OSD-CBCT. They focus on aspects such as noise level, contrast, and CT numbers. The structural similarity index measure (SSIM)  \cite{ssim}assesses the structural congruence between generated and true OSD-CBCTs, reflecting anatomical fidelity. The normalized cross-correlation coefficient (NCC) quantifies the statistical correlation, emphasizing pattern alignment between the generated and true images. Additionally, the Fréchet inception distance (FID) \cite{FID} and the learned perceptual image patch similarity (LPIPS) \cite{LPIPS} metrics appraise the visual realism of the generated images from a statistical perspective. These latter metrics compute the statistical distributions of features or activations derived from deep learning models, extracted from both true and generated images. This provides insights into their perceptual realism. For evaluation purposes, lower values of MAE, FID, and LPIPS indicate better performance. Higher values of PSNR, SSIM, and NCC signify superior reconstruction quality.
	
	In addition to global evaluations, our study incorporates localized, ROI-based assessments for organs and GTV regions using the same metrics. This method employs manual tumor and organ delineation masks from truth OSD-CBCTs. These masks encompass GTV and other vital structures such as the esophagus, heart, lung, and spinal cord. By extracting ROIs from both true and generated OSD-CBCT, our evaluation uses the aforementioned metrics to quantify the accuracy within each designated ROI. This ensures a detailed and precise analysis of image performance in clinically significant structures.
	
	\subsubsection{Analysis 1: OSD-CBCT Quality Analysis Results}
	The averaged quantitative evaluation results for the whole body OSD-CBCTs are reported in Table 1. Our method achieves high image quality in both imaging scenarios, with direct and indirect single-angle projection guidance. Specifically, the PC-DDPM achieves a PSNR of 37.42 dB and an SSIM of 0.97 for the first-view OSD-CBCT with direct projection guidance, and a PSNR of 36.76 dB and an SSIM of 0.96 for the subsequent-view OSD-CBCT with indirect projection guidance (without a same phase projection). These values demonstrate the decent OSD-CBCT fidelity. Additionally, the MAE is 19.96 HU with projection and 23.65 HU without projection, indicating precise generation accuracy. Visual evaluation metrics further support these findings, with FID scores of 0.33 and 0.34, and LPIPS scores of 0.09 and 0.11, respectively. These metrics highlight the visual quality and perceptual similarity of the generated OSD-CBCT images.
	
	For the GTV, our method also shows high image quality. The OSD-CBCTs achieves a PSNR of 33.14 dB and an SSIM of 0.93 with direct projection guidance, and a PSNR of 32.79 dB and an SSIM of 0.93 with indirect projection guidance. The MAE is 39.75 HU and 43.02 HU for two imaging scenarios, respectively, reflecting the model's accuracy in tumor region generation. Visual metrics further confirm the performance, with FID scores of 0.61 with direct projection and 0.65 with indirect projection, and LPIPS scores of 0.32 and 0.34, respectively. These results emphasize the robustness of our method in accurately displaying tumors, crucial for visually tumor tracking and dose calculation in treatment planning.

	\begin{table*}[!t]
		\centering
		\resizebox{\columnwidth}{!}{%
			\begin{tabular}{c c c c c c c c}
				\hline
				\multirow{2}{*}{\textbf{ }} & \multirow{2}{*}{\textbf{ }} & \multirow{2}{*}{\textbf{MAE (HU) (↓)}} & \multirow{2}{*}{\textbf{PSNR (dB) (↑)}} & \multirow{2}{*}{\textbf{SSIM (↑)}} & \multirow{2}{*}{\textbf{NCC (↑)}} & \multirow{2}{*}{\textbf{FID (↓)}} & \multirow{2}{*}{\textbf{LPIPS (↓)}} \\
				& & & & & & & \\
				\hline
				\multirow{2}{*}{\textbf{Whole Volume}} & With projection & 19.96±5.29 & 37.42±1.58 & 0.97±0.02 & 1.00±0.00 & 0.33±0.12 & 0.09±0.06 \\
				& Without projection & 23.65±4.03 & 36.76±1.18 & 0.96±0.02 & 1.00±0.00 & 0.34±0.12 & 0.11±0.06 \\
				\hdashline
				\multirow{2}{*}{\textbf{GTV}} & With projection & 39.75±7.39 & 35.14±1.06 & 0.93±0.02 & 0.96±0.02 & 0.61±0.13 & 0.32±0.12 \\
				& Without projection & 43.02±10.19 & 34.79±1.34 & 0.93±0.02 & 0.95±0.02 & 0.65±0.04 & 0.34±0.09 \\
				\hdashline
				\multirow{2}{*}{\textbf{Esophagus}} & With projection & 38.56±5.44 & 33.22±1.24 & 0.42±0.03 & 0.99±0.01 & 0.64±0.05 & 0.09±0.06 \\
				& Without projection & 39.60±4.47 & 32.99±0.94 & 0.94±0.03 & 0.99±0.01 & 0.46±0.15 & 0.19±0.10 \\
				\hdashline
				\multirow{2}{*}{\textbf{Heart}} & With projection & 39.15±5.88 & 33.13±1.02 & 0.95±0.03 & 0.99±0.00 & 0.56±0.06 & 0.17±0.05 \\
				& Without projection & 40.17±5.25 & 32.92±0.83 & 0.94±0.03 & 0.99±0.00 & 0.61±0.08 & 0.17±0.06 \\
				\hdashline
				\multirow{2}{*}{\textbf{Lung}} & With projection & 37.18±7.49 & 32.66±1.42 & 0.95±0.03 & 1.00±0.00 & 0.25±0.05 & 0.07±0.04 \\
				& Without projection & 38.29±6.69 & 32.42±1.16 & 0.94±0.03 & 1.00±0.00 & 0.26±0.10 & 0.07±0.05 \\
				\hdashline
				\multirow{2}{*}{\textbf{Spinal cord}} & With projection & 43.67±8.08 & 31.70±1.40 & 0.94±0.03 & 0.96±0.02 & 0.46±0.16 & 0.18±0.06 \\
				& Without projection & 46.25±6.91 & 31.24±1.11 & 0.94±0.03 & 0.96±0.02 & 0.44±0.22 & 0.18±0.07 \\
				\hline
			\end{tabular}%
		}
		\caption{Quantitative evaluation of generated first- (directly guided by first-view X-ray projection) and subsequent-view (indirectly guided by first-view X-ray projection) OSD-CBCT in terms of MAE, PSNR, SSIM, NCC, FID, and LPIPS on the whole body and five different ROIs, including gross tumor volume (GTV), esophagus, heart, lung, and spinal cord.}
	\end{table*}

	\subsection{Evaluation 2: OSD-CBCT’s Tumor Motion Analysis}
	To evaluate the OSD-CBCT’s effectiveness on tumor motion tracking, a key application of our A-SI framework, certified medical physicists used RayStation 2023B to manually contour the GTV on both true and generated first- and subsequent-view OSD-CBCTs. We calculated the GTV center of mass (COM) distances between the ground truth and generated OSD-CBCTs to assess tumor motion errors, where lower values indicate better accuracy. To further evaluate tumor motion, we measured the maximum tumor motion range across 10 phases for each patient. We normalized the GTV COM distance by this range to obtain the relative GTV COM distance, which reflects the relationship between tumor motion error and the maximum patient tumor motion, with lower values indicating better performance. We also evaluated tumor similarity using the Dice Score Coefficient (DSC) for overall structure of the generated and truth tumors and reported the 95\% Hausdorff Distance (HD95) to quantify how closely the generated tumors’ boundaries align with the truth tumor’s boundaries. 
	
	\subsubsection{Analysis 2: OSD-CBCT’s Tumor Motion Analysis results}
	Fig. 3A and B present the target CBCT, first-view OSD-CBCT (directly guided by first-view X-ray projection) and subsequent-view OSD-CBCT (directly guided by first-view X-ray projection), and their corresponding tumors, which delineated by certified medical physicists in RayStation 2023B. The OSD-CBCTs present tumors can be clearly delineated by medical physicists and located in the accurate locations. Table 2 presents GTV COM distances, relative GTV COM distances, DSC, and HD95 for the tumors from both the CBCT and OSD-CBCT across all 22 patients. For CBCT tumor tracking, the first-view OSD-CBCT achieves a COM distance of 0.50 ± 0.28 mm, relative COM distance of 0.06 ± 0.03, HD95 of 1.86 ± 0.58 mm, and a DSC of 0.87 ± 0.04. The subsequent-view OSD-CBCT yields a COM distance of 0.56 ± 0.39 mm, relative COM distance of 0.07 ± 0.04, HD95 of 1.96 ± 0.58 mm, and a DSC of 0.85 ± 0.05. The OSD-CBCTs demonstrate accurate tumor appearance and location across all metrics, enabling tumor motion tracking during the therapy.
	
	\begin{figure}
		\centering
		\noindent \includegraphics*[width=6.50in, height=4.20in, keepaspectratio=true]{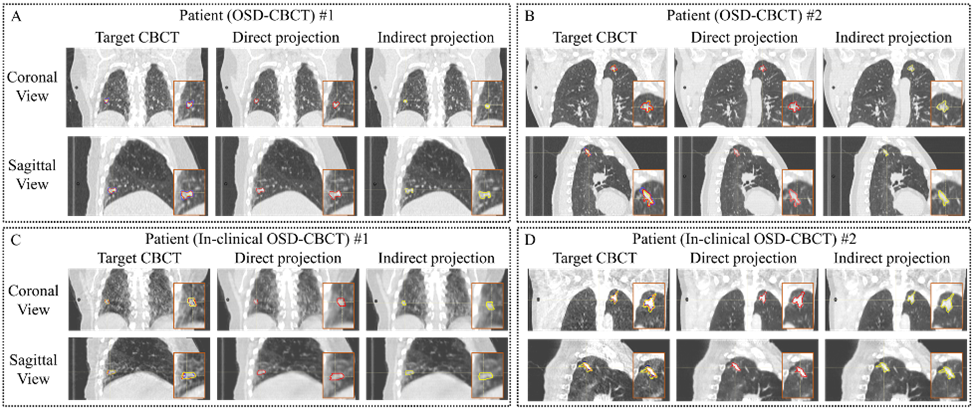}
		
		\noindent Figure 3: Visual examples of OSD-CBCT for first view (directly guided by first-view X-ray projection) and subsequent-view (indirectly guided by first-view X-ray projection) in RayStation 2023B. The top two rows show results using surface image and CBCT’s X-ray projection simulated by 4DCTs, and the bottom two rows show results using in-clinical CBCT’s surface image and projections (two patients each). In each block, the first row displays the coronal view of the ground truth (1st column), first-view OSD-CBCT (2nd column), and subsequent-view OSD-CBCT (3rd column). The second row presents the sagittal views of the corresponding OSD-CBCT images. Orange boxes highlight zoomed-in tumor regions, where blue contours represent tumors in the true OSD-CBCT, red contours indicate tumors in the generated first-view OSD-CBCT, and yellow contours show tumors in the generated subsequent-view OSD-CBCT. Window level is [-600, 1000] HU.
	\end{figure}

	Other organs, such as the esophagus, heart, lung, and spinal cord, also maintain high quality from our PC-DDPM method. For instance, in the lung region, the method achieves PSNR values of 33.66 with direct projection and 33.42 with indirect projection, and SSIM values of 0.95 and 0.94, respectively. The visual metrics for the lung are also favorable, with FID scores of 0.25 with projection and 0.26 without projection, and LPIPS scores of 0.07 in both scenarios. These results consistently demonstrate that our PC-DDPM model can effectively generate high quality OSD-CBCT containing various clinically significant structures in radiotherapy, regardless of the presence of projection data.
	
	\begin{table*}[!t]
		\centering
		\resizebox{\columnwidth}{!}{%
			\begin{tabular}{c c c c c c}
				\hline
				\multirow{2}{*}{\textbf{ }} & \multirow{2}{*}{\textbf{ }} & \multirow{2}{*}{\textbf{COM Distance (mm)}} & \multirow{2}{*}{\textbf{Relative COM Distance}} & \multirow{2}{*}{\textbf{DSC}} & \multirow{2}{*}{\textbf{HD95 (mm)}}  \\
				& & & & & \\
				\hline
				\multirow{2}{*}{\textbf{OSD-CBCT}} &Direct projection guided & 0.50±0.28 & 0.06±0.03 & 0.87±0.04 & 1.86±0.58 \\
				& Indirect projection guided & 0.56±0.39 & 0.07±0.04 & 0.85±0.05 & 1.96±0.58 \\
				\hdashline
				\multirow{2}{*}{\textbf{In-clinical OSD-CBCT}} & Direct projection guided & 0.52±0.22 & N/A &0.88±0.01 & 1.89±0.12 \\
				& Indirect projection guided & 0.60±0.25 & N/A & 0.87±0.01 & 2.01±0.19 \\
				\hline
			\end{tabular}%
		}
		\caption{Tumor motion tracking evaluation in first- (directly guided by first-view X-ray projection) and subsequent-view (indirectly guided by first-view X-ray projection) OSD-CBCT and in-clinical OSD-CBCT. The comparisons included GTV COM distances, relative GTV COM distances, DSC, HD95 for the tumor in ground truth and OSD-CBCT. The relative COM distance is not applicable for in-clinical CBCT.}
	\end{table*}
	
	\subsection{Evaluation 3: Projection Angle Analysis for First-view OSD-CBCT with Projection}
	During radiotherapy, the CBCT imager rotates with Linac delivery gantry. In our A-SI framework, projections from different radiation windows may come from varying angles due to this rotation. Therefore, the proposed PC-DDPM must generate consistent high-quality first-view OSD-CBCT under supervision from projections at different angles. To validate this, we evaluated PC-DDPM’s performance using projections from five specific angles:  \(0^\circ\),\(30^\circ\), \(45^\circ\), \(60^\circ\), and \(90^\circ\). The evaluation included MAE and LPIPS metrics, with both global and ROI-based tumor assessments for comprehensive analysis.
	
	\subsubsection{Analysis 3: Projection Angle Analysis Results}
	Fig. 4 illustrates the quantitative evaluation of the first-view OSD-CBCT across different single-angle projection angles (\(0^\circ\),\(30^\circ\), \(45^\circ\), \(60^\circ\), and \(90^\circ\)) for the whole body and GTV. For the whole body, MAE values remain consistently low, ranging from 19.71 HU at 0 degrees to 20.20 HU at 90 degrees, indicating high generation accuracy. LPIPS values are stable as 0.10, suggesting consistent visual quality. For the GTV, MAE values range from 39.10 HU to 41.15 HU, demonstrating robust tumor generation. LPIPS values for the GTV fluctuate minimally, between 0.31 and 0.33. These results show that although the OSD-CBCT generated quality decreases with the change in projection angle from 0 to 90 degrees, it still performs well across various projection angles. The global and local MAE (image accuracy) and LPIPS (visual realism) values remain relatively stable for both the whole body and GTV. This demonstrates that the PC-DDPM is applicable for the proposed A-SI framework when the single-angle projection angle changes during collection.
	\begin{figure}
		\centering
		\noindent \includegraphics*[width=6.50in, height=4.20in, keepaspectratio=true]{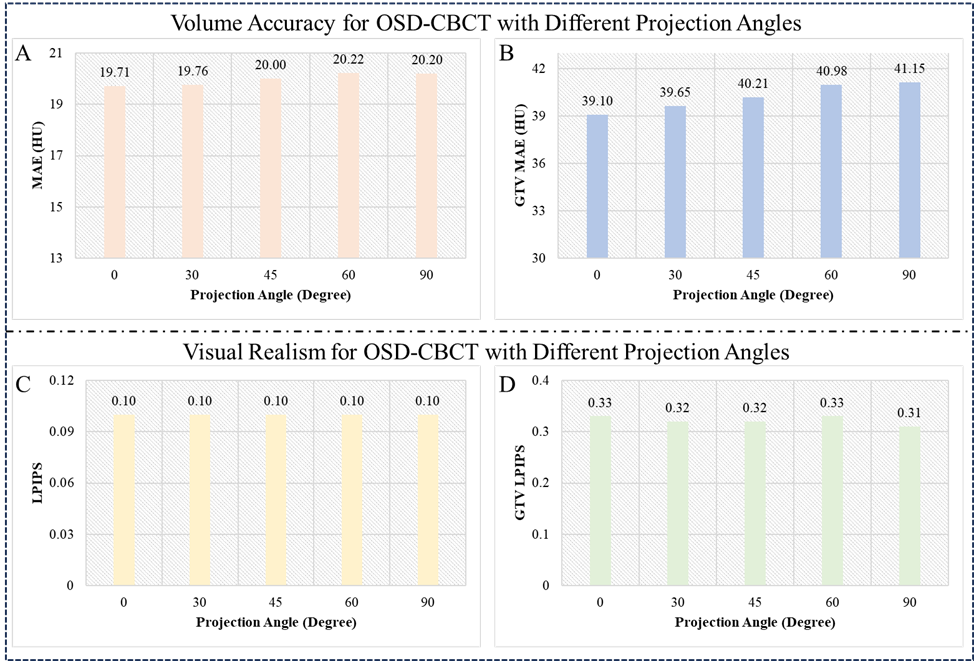}
		
		\noindent Figure 4: First-view OSD-CBCT (directly guided by first-view X-ray projection) accuracy and realism with single-angle projections collected at different angles(\(0^\circ\),\(30^\circ\), \(45^\circ\), \(60^\circ\), and \(90^\circ\)). A and C show whole-body MAE and LPIPS, while B and D display tumor-specific MAE and LPIPS. Consistent performance across angles indicates improved tumor motion tracking in the A-SI framework.
	\end{figure}

	\subsection{Evaluation 4: Projection Collection Frequency Analysis for Subsequent-view OSD-CBCT without Projection}
	In a radiation window of the A-SI framework, the subsequent-view OSD-CBCT, without projections, is guided by the first-view OSD-CBCT, which includes projection data. The radiation window length can be prolonged such that it collects less X-ray projections to reduce radiation dose, but it also means more views without the corresponding X-ray projections projection. The PC-DDPM uses the first-view OSD-CBCT to guide multiple subsequent views, but a larger time gap between the first- and current-view OSD-CBCTs can reduce its reliability due to patient motion and tumor changes, increasing discrepancies. We aimed to assess how projection collection frequency affects OSD-CBCT quality and identify the optimal frequency for clinical use. We tested four different frequencies: one projection for every three, five, seven, and ten surface images (default). In other words, we used the OSD-CBCT at these intervals with the inferred subsequent-view OSD-CBCT, rather than the validation OSD-CBCT. MAE and LPIPS metrics were used for both global and tumor-specific ROI assessments to ensure a thorough evaluation.
	
	\subsubsection{Analysis 4: Projection Collection Frequency Analysis Results}
	Fig. 5 shows the quantitative evaluation of the subsequent-view OSD-CBCT across different projection collection frequencies (3, 5, 7, and 10 surface images per projection) for the whole body and GTV. For the whole body, MAE values range from 22.93 HU to 24.15 HU, and LPIPS values remain constant at 0.11, indicating consistent accuracy and visual quality. For the GTV, MAE values range from 42.87 HU to 45.08 HU, and LPIPS values vary slightly from 0.33 to 0.34, demonstrating robust performance. These results confirm that our method maintains high quality and accuracy even when collecting only one projection per ten surface images. The ability of the PC-DDPM to generate OSD-CBCT in low collection frequency is preferable by minimizing radiation dose exposure to the patient. 
	
	\begin{figure}
		\centering
		\noindent \includegraphics*[width=6.50in, height=4.20in, keepaspectratio=true]{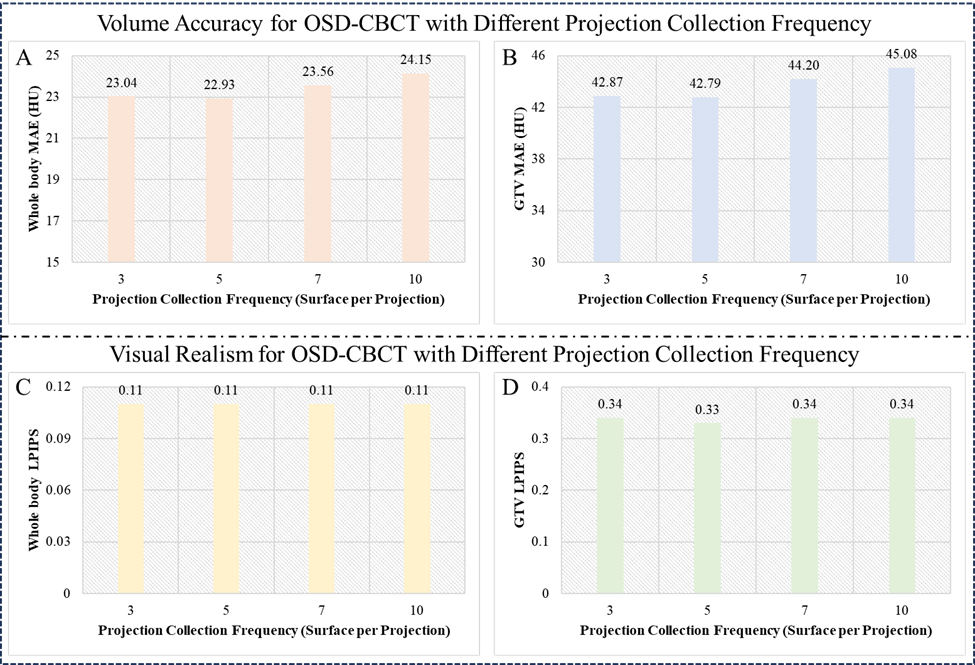}
		
		\noindent Figure 5: Subsequent-view OSD-CBCT (without projection so indirectly guided by first-view X-ray projection) accuracy and realism for single-angle projections collected at different frequencies (3, 5, 7, 10 surface image per projection). A and C present whole-body MAE and LPIPS, while B and D show tumor-specific MAE and LPIPS. Strong performance at lower projection frequencies indicates effective tumor tracking in the A-SI framework with reduced radiation exposure. 
	\end{figure}
	
	\subsection{Evaluation 5: OSD-CBCT Efficiency Analysis}
	This evaluation analyzes the generative speed of the PC-DDPM and its impact on OSD-CBCT quality. Faster generation enhances the A-SI framework's potential for real-time tumor motion tracking, but the framework must still maintain accurate OSD-CBCT. The speed and quality depend on the number of generation timesteps, with generation time increasing linearly with timesteps. Fewer timesteps can reduce generation time but may compromise OSD-CBCT quality. We report the generation time and OSD-CBCT quality for PC-DDPM with 5, 10 (default), 25, 50, and 100 timesteps. MAE and LPIPS metrics, along with global and tumor-specific ROI assessments, are included in the quantitative analysis.
	
	\subsubsection{Analysis 5: OSD-CBCT Efficiency Analysis Results}
	Fig. 6 shows a visual example and Fig. 7 evaluates the quality of the first- and subsequent-view OSD-CBCT generated using different generation timesteps (G = 5, 10, 25, 50, 100). In terms of MAE and LPIPS for both the whole body and GTV, at timestep 5, the results show high MAE and LPIPS values, indicating poor global and local image quality for both imaging scenarios. However, at timestep 10 (default setting), there is a significant improvement. For the first-view OSD-CBCT, the whole-body MAE was reduced to 19.92 HU and GTV MAE to 39.75 HU. The LPIPS values drop to 0.09 and 0.11, respectively. The subsequent-view OSD-CBCT also demonstrates the same MAE and LPIPS trends. Increasing the timesteps beyond 10 results in only marginal improvements in both MAE and LPIPS, while significantly increasing the generation time. To minimize generation time and push towards real-time application, timestep 10 appears optimal, balancing performance and efficiency.
	
	\begin{figure}
		\centering
		\noindent \includegraphics*[width=6.50in, height=4.20in, keepaspectratio=true]{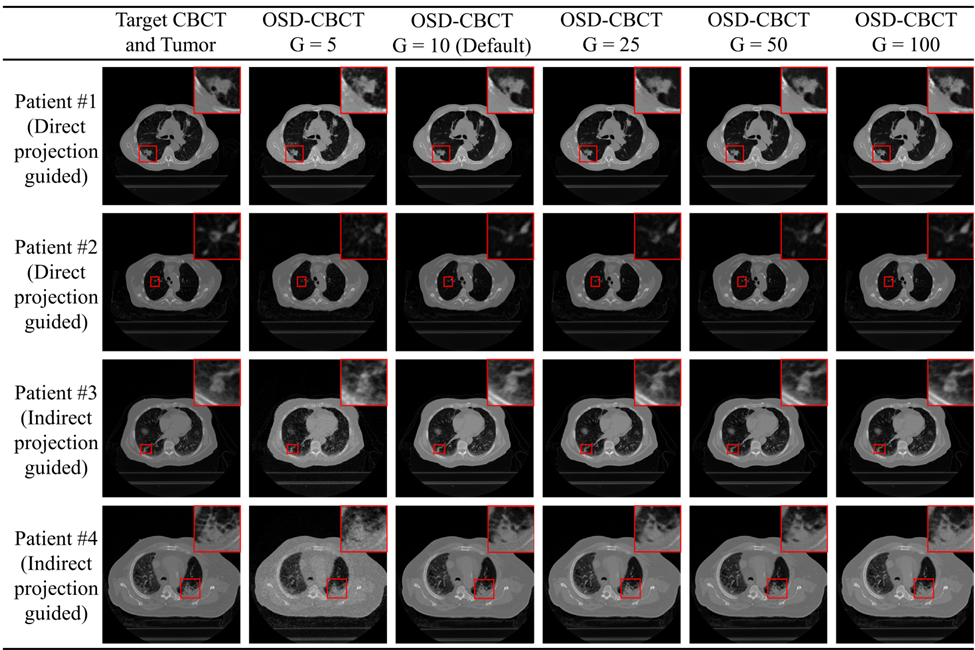}
		
		\noindent Figure 6: Four visual examples of OSD-CBCT and zoomed-in tumor regions (red boxes) generated by PC-DDPM with varying generation timesteps. The first two rows are generated with X-ray projection (directly guided by first-view X-ray projection), while the last two rows are without projection (indirectly guided by projection). The first column shows the target OSD-CBCT, and the second to sixth columns display generated OSD-CBCTs with different timesteps (G = 5, 10, 25, 50, 100).
	\end{figure}

	\begin{figure}
		\centering
		\noindent \includegraphics*[width=6.50in, height=4.20in, keepaspectratio=true]{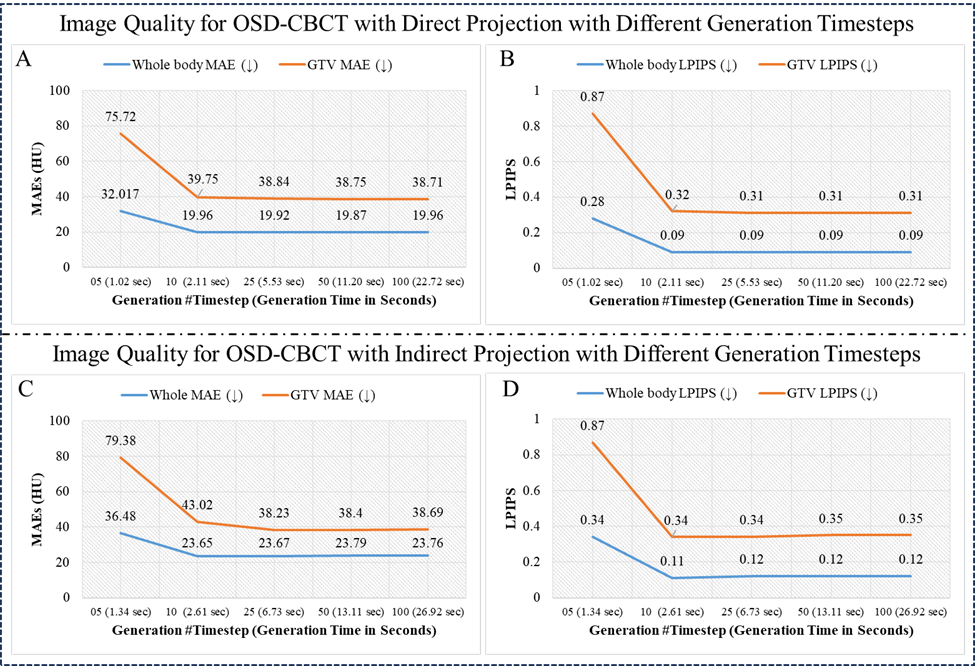}
		
		\noindent Figure 7: OSD-CBCT accuracy and realism for different generation timesteps and the corresponding generation time. A and B show whole-body and GTV MAE and LPIPS for first-view OSD-CBCT with projection (directly guided by first-view X-ray projection). C and D present whole-body and GTV MAE and LPIPS for subsequent-view OSD-CBCT without projection (indirectly guided by projection).
	\end{figure}
	
	\subsection{Evaluation 6: PC-DDPM Refinement Strategy Analysis}
	The PI and Cycle-C refinement strategies are key to high-quality first- and subsequent-view OSD-CBCT generation, directly or indirectly guided by first-view X-ray projection, respectively. To assess the PI refinement's contribution, we generated first-view OSD-CBCT using only the CBCT-DDPM, excluding the projection-DDPM and GTM, based on surface images and single-angle projections. To evaluate Cycle-C refinement, we generated subsequent-view OSD-CBCT only using the surface image, without projection and the referenced first-view OSD-CBCT, excluding the cycle-consistency mechanism from PC-DDPM. Performance was measured using global and ROI-based tumor quality metrics. Additionally, a Student’s T-test was applied to statistically compare OSD-CBCT results with and without each refinement, using a P-value threshold of 0.05 for significance.
	
	\subsubsection{Analysis 6: PC-DDPM Refinement Strategy Analysis Results}
	The results in Table 3 show that both PI and Cycle-C refinement strategies significantly improve OSD-CBCT quality across most metrics. For whole-body evaluations, PI refinement outperforms the no-refinement approach, with lower MAE (19.96 vs. 29.38 HU), higher PSNR (37.42 vs. 34.90 dB), and higher SSIM (0.97 vs. 0.93), all with P-values <0.05. Similarly, Cycle-C refinement yields better results than no-refinement, with lower MAE (23.65 HU vs. 33.24 HU) and higher PSNR (36.76 dB vs. 32.43), with P-values <0.05. In GTV assessments, both PI and Cycle-C refinements show significant improvements in MAE, PSNR, SSIM, and NCC, with all P-values <0.05, demonstrating their effectiveness in generating high quality OSD-CBCTs. In addition, the refinement strategies have minor effect on the FID and LPIPS.
	
	\begin{table*}[!t]
		\centering
		\resizebox{\columnwidth}{!}{%
			\begin{tabular}{c c c c c c c c}
				\hline
				\multirow{2}{*}{\textbf{ }} & \multirow{2}{*}{\textbf{ }} & \multirow{2}{*}{\textbf{MAE (HU) (↓)}} & \multirow{2}{*}{\textbf{PSNR (dB) (↑)}} & \multirow{2}{*}{\textbf{SSIM (↑)}} & \multirow{2}{*}{\textbf{NCC (↑)}} & \multirow{2}{*}{\textbf{FID (↓)}} & \multirow{2}{*}{\textbf{LPIPS (↓)}} \\
				& & & & & & & \\
				\hline
				\multirow{2}{*}{\textbf{Whole Body}} & PI & 19.96±5.29* & 37.42±1.58* & 0.97±0.02* & 1.00±0.00 & 0.33±0.12 & 0.09±0.06  \\
				& No refinement & 29.38±5.98 & 34.90±1.87 & 0.93±0.03 & 1.00±0.00 & 0.33±0.10 & 0.09±0.08\\
				& P-value & $<$0.05 & $<$0.05 & $<$0.05 & 1.00 & 0.88 & 0.95 \\
				\hline
				\multirow{2}{*}{\textbf{Whole Body}} & Cycle-C & 23.65±4.03* & 36.76±1.18* & 0.96±0.02* & 1.00±0.00 & 0.34±0.12 & 0.11±0.06 \\
				& No refinement & 33.24±7.30 & 32.43±1.91 & 0.92±0.02 & 1.00±0.00 & 0.35±0.09 & 0.11±0.08 \\
				& P-value & $<$0.05 & $<$0.05 & $<$0.05 & 1.00 & 0.52 & 0.87 \\
				\hline
				\multirow{2}{*}{\textbf{GTV}} & PI & 39.75±7.39* & 33.14±1.06* & 0.93±0.02* & 0.96±0.02* & 0.61±0.13 & 0.32±0.12 \\
				& No refinement & 45.63±15.92 & 34.11±2.00 & 0.90±0.04 & 0.92±0.03 & 0.60±0.11 & 0.32±0.10 \\
				& P-value & $<$0.05 & $<$0.05 & $<$0.05 & $<$0.05 & 0.91 & 0.97 \\
				\hline
				\multirow{2}{*}{\textbf{GTV}} & Cycle-C & 43.02±10.19* & 32.79±1.34* & 0.93±0.02* & 0.95±0.02* & 0.65±0.04 & 0.34±0.09 \\
				& No refinement & 48.83±29.60 & 32.50±5.04 & 0.89±0.05 & 0.87±0.15 & 0.67±0.15 & 0.34±0.06\\
				& P-value & $<$0.05 & $<$0.05 & $<$0.05 & $<$0.05 & 0.14 & 0.90 \\
				\hline
			\end{tabular}%
		}
		\caption{Quantitative results comparing first- and subsequent-view OSD-CBCTs with and without PI refinement and Cycle-C refinement. P-values from the Student’s T-test are shown for each comparison. The best results are bolded, and results with statistical significance are marked with a star (*).}
	\end{table*}
		
	\subsection{Evaluation 7: In-clinical OSD-CBCT Analysis}
	We aim to validate the proposed PC-DDPM using real-world clinical CBCT data in addition to CBCTs simulated from 4DCTs. Unlike our previous analysis, which used different phases of 4DCTs for both training and testing to generate OSD-CBCT, this analysis trains the model on 4DCT data and tests its ability to generate OSD-CBCTs from real clinical CBCT data. The goal is twofold: first, to demonstrate that PC-DDPM can generate OSD-CBCTs out-of-distribution of the training data. Secondly, we show that the generated OSD-CBCTs have correct tumor motion of the real clinical CBCTs, supporting their potential for clinical application. We collected 4DCT and in-clinical CBCT data from the treatment days of three patients. All phases of 4DCT were used for training, and the CBCT was used for testing. For first-view OSD-CBCT generation directly guided by first-view X-ray projection, the model used X-ray projection and simulated surface images from the CBCT. For subsequent-view OSD-CBCT generation indirectly guided by projection, we used the 90\% phase training 4DCT surface image and projection to generate a reference OSD-CBCT. This reference guided in-clinical OSD-CBCT generation from real CBCT surface images through Cycle-C refinement. We evaluated the in-clinical OSD-CBCTs using tumor tracking metrics to ensure accuracy and reliability in clinical use.
	
	\subsubsection{Analysis 7: In-clinical OSD-CBCT Analysis results}
	Fig. 3C and D show in-clinical CBCT, first-view in-clinical OSD-CBCT (directly guided by first-view X-ray projection), subsequent-view in-clinical OSD-CBCT (indirectly guided by first-view X-ray projection), and their corresponding tumors. The OSD-CBCTs, generated using the surface image and single-angle projections from in-clinical CBCT, clearly delineate tumors, accurately located for medical assessment. Table 2 summarizes the GTV COM distances, relative COM distances, DSC, and HD95 for tumors in both in-clinical CBCT and OSD-CBCT across all 3 patients. For tumor tracking in in-clinical CBCT, the OSD-CBCT with projection achieved a COM distance of 0.52 ± 0.22 mm, HD95 of 1.89 ± 0.12 mm, and DSC of 0.88 ± 0.01. The without projection OSD-CBCT showed a COM distance of 0.60 ± 0.25 mm, HD95 of 2.01 ± 0.19 mm, and DSC of 0.87 ± 0.01. These results demonstrate that in-clinical OSD-CBCT provides accurate tumor tracking, indicating potential for use in real-world therapy.

	\section{Discussion}
	We propose an advanced imaging framework for surface-guided radiotherapy, called A-SI, designed for real-time tumor tracking while minimizing radiation exposure. A-SI uses 4DCT and surface images collected at the treatment planning stage to train a patient-specific diffusion generative model, generating CBCT-like volumetric images called OSD-CBCTs. During radiation delivery, A-SI divides the process into multiple windows, each covering a range of angles. In each window, an optical surface camera continuously captures real-time 2D surface images, while a CBCT system takes a single-angle cone beam X-ray projection at the first phase (first view). The patient-specific model uses the surface image and the corresponding projection to generate the first-view OSD-CBCT. For subsequent views, the model generates OSD-CBCT from surface images, along with the first-view OSD-CBCT as a reference. The radiation linear accelerator then rotates to the next window, continuing the process of collecting surface images and single-angle projections, so keep generating real-time OSD-CBCT volumes. 
	To implement A-SI, we propose the PC-DDPM as the generative model. PC-DDPM leverages the DDPM’s ability to iteratively refine Gaussian noise into visually realistic images. We introduce two key innovations for two imaging scenarios: first-view OSD-CBCT generation with projection, and subsequent-view OSD-CBCT generation without projection. For the first scenario, we propose the PI refinement strategy to guide the OSD-CBCT generation by the projection directly. This uses dual-domain DDPMs: the CBCT-DDPM generates OSD-CBCT from surface images, while the Projection-DDPM generates full-angle projections from a single-angle projection. These outputs guide each other iteratively through a GTM to create a physics-integrated OSD-CBCT. For the second scenario, we propose the Cycle-C refinement strategy to use the first-view OSD-CBCT to guide the generation of subsequent-view OSD-CBCTs. Following the cycle-consistency principle, the current subsequent-view OSD-CBCT should regenerate the first-view OSD-CBCT. The difference between the regenerated and original first-view OSD-CBCT is used to refine the subsequent-view OSD-CBCT generation. These refinement strategies enable the DDPM to flexibly accept both surface images and single-angle projections or surface images alone, while ensuring the generated OSD-CBCT adheres to real-world cone beam geometry for precise radiation delivery.
	To validate the PC-DDPM and its role in the A-SI framework, we conducted a comprehensive simulation study using 4DCT data from 22 patients to simulate target CBCT to train the model. The model's performance was assessed qualitatively through visual examples and quantitatively using metrics such as MAE, PSNR, SSIM, and NCC. Additionally, visual appearance was evaluated with FID and LPIPS. Global assessments covered the entire OSD-CBCT volume, while local evaluations focused on five ROIs: GTV, esophagus, heart, lung, and spinal cord. Qualitatively, both first- and subsequent-view OSD-CBCTs preserved realistic anatomical and textural details, particularly for the whole volume and GTV, which is critical for accurate tumor identification and precise radiation delivery. Quantitative evaluations of other ROIs also demonstrated highly accurate and realistic results, enhancing treatment precision while minimizing patient dose exposure and side effects.
	More importantly, we validated the ability of OSD-CBCT to accurately present tumors with correct appearance and location, demonstrating the effectiveness of the A-SI framework for tumor motion tracking. As shown in Fig. 4 and Table 2, the average GTV COM distance is 0.50 mm for first view and 0.56 mm for subsequent-view OSD-CBCT across all 22 patients, well below the clinically acceptable motion error margin of ~2 mm. The average DSC and HD95 values indicate substantial similarity in location, shape, and boundary between the OSD-CBCT and ground truth tumors. This confirms the accuracy of tumor representation in both imaging scenarios, enabling the A-SI framework where only the first view uses projection and subsequent views do not. Moreover, OSD-CBCT generated continuously throughout therapy maintains high tumor location accuracy, supporting real-time tumor tracking.
	Both qualitative and quantitative results show that OSD-CBCT achieves accurate tumor appearance and location, both with and without projection. The first-view OSD-CBCT uses the patient's specific anatomy and a single-angle projection to place the tumor correctly. It works because the single-angle projection depicts the tumor along the z-axis, where tumor motion is usually more significant than along the other axes. As a result, the first-view OSD-CBCT can achieve high tumor accuracy. In the subsequent views, OSD-CBCT uses the first view as a reference to position the tumor correctly, even without additional projection. This is possible because 4DCT from training data provides information about how the tumor moves across from the first view to the subsequent views. Therefore, the generation of subsequent-view OSD-CBCT is more a patient-specific tumor motion estimation than a simple image synthesis. The use of the first-view as a reference in PC-DDPM is critical for accurate tumor location in subsequent views.
	This framework generates synthetic CT images with minimal radiation, providing detailed anatomical information essential for lesion localization. It has the potential to enable real-time image-guided radiation therapy. Each patient receives a unique model, trained individually. Training occurs during the treatment planning stage, running in the background. Once trained, the model works with surface imaging devices to predict 3D patient anatomy in real-time. This improves patient setup, verification, and online monitoring for both photon and proton therapy. The framework can also support real-time imaging for radiosurgery, interventional procedures, and ultra-high dose rate FLASH radiotherapy. However, we should notice that the relative COM distance, around 0.06, shows that the framework maintains high accuracy during typical patient motion (5 mm to 15 mm). On the contrary, in rare cases of extreme motion, accuracy may be limited. Future work will focus on improving the A-SI framework for these scenarios.
	We also applied in-clinical CBCT, which are additional to the 4DCTs, to simulate surface images and X-ray projections to generate in-clinical OSD-CBCT and evaluated its tumor motion accuracy. The first- and subsequent-view in-clinical OSD-CBCT showed low COM distance (~0.60 mm), high DSC (~0.87), and low HD95 (~2.00 mm). These results confirm that the PC-DDPM method can reliably generate OSD-CBCT with accurate tumor representation, even when using real-world CBCT surface images and X-ray projections. This demonstrates its strong potential for tumor tracking in clinical radiotherapy settings.
	For first-view OSD-CBCTs with projection, a single-angle projection angle evaluation validated that the PC-DDPM can accept projections from different angles to generate accurate OSD-CBCTs. During radiation delivery, the radiation linear accelerator and CBCT imager rotate, resulting in projections collected from various angles in different delivery windows. This evaluation demonstrated that the PC-DDPM can generate high-quality OSD-CBCTs regardless of the projection angle, proving its applicability to generate high quality OSD-CBCT during the whole delivery. For the subsequent-view CBCT without projection, which uses the first-view OSD-CBCT as a reference, a projection collection frequency evaluation was performed. Although the PC-DDPM is flexible enough to use the first-view OSD-CBCT to guide any subsequent-view OSD-CBCT (e.g., OSD-CBCT from 10 or 20 phases later), the patient's natural movements during delivery reduce the reliability of first-view OSD-CBCT for the subsequent phase far from the starting phase. Dividing the delivery into smaller windows allows for collecting higher frequency projections, potentially achieving higher accuracy. However, higher frequency projections increase radiation exposure. This evaluation aimed to find an optimal projection collection frequency to obtain sufficient quality OSD-CBCT while minimizing radiation exposure. The results showed that collecting one projection for every ten surface images provided good quality OSD-CBCTs, making it a reasonable frequency. Future clinical evaluations will aim to determine the quality of OSD-CBCTs with more collection frequency choices, including lower frequencies, to find the optimal projection frequency for clinical use.
	Overall, global and local evaluations demonstrated high image accuracy and visual realism of the generated CBCTs. Angle and projection evaluations confirmed the applicability of the PC-DDPM in the A-SI framework. Additionally, an ablation study demonstrated that the PI and Cycle-C refinement strategies significantly improved generation accuracy, but not visual realism compared to the PC-DDPM without these refinements, as evidenced by both quantitative and statistical results. However, we should consider that the accuracy benefits arise from the A-SI framework design: using low-frequency single-angle projections to assist the surface-to-OSD-CBCT generation process. The refinement strategies are designed to utilize these projections to enable the A-SI framework. This reveals that it is possible to design other potential mechanisms if they can utilize the projection to assist OSD-CBCT generation from surface images, motivating further research into more effective and efficient methods to enable the A-SI framework.
	
	Furthermore, an efficiency evaluation showed that 10 generation timesteps, as shown in Fig. 7, taking about 2.5 seconds on an A100 GPU to generate a 512×512×128 volume, provided an optimal balance between image quality and generation time. Fewer timesteps resulted in noisier images, potentially hindering tumor identification. More timesteps slightly improved image quality but significantly increased generation time. Despite the benefits of OSD-CBCT using the proposed 10 timesteps for tumor tracking and the validation of its applicability with PC-DDPM in the proposed A-SI framework, there is a need for improvement. The primary concern is this relative inefficiency of PC-DDPM, which limits its use for real-time tumor tracking (e.g., 1 second for OSD-CBCT generation). Enhancing PC-DDPM's efficiency, potentially by integrating high-efficiency DDPMs such as the One-step Consistency Model  \cite{consistencymodel} and DiffuseVAE \cite{diffusevae}, is a pertinent area for future research to achieve real-time volumetric imaging during radiation delivery. Additionally, using lightweight networks to learn the generation processes in PCDDPM could improve generation speed, pushing it closer to real-time capability. A more efficient design for dual-domain DDPMs could also enhance generation speed while maintaining OSD-CBCT quality. Beyond efficiency improvements, further experiments with a larger patient cohort are necessary to ensure the method's generalizability across diverse patient profiles. Applying A-SI to real clinical therapy to validate its utility in radiotherapy is a significant and exciting direction for our follow-up research.
	
	\section{Conclusion}
	We propose the A-SI framework, a new method to generate real-time, radiation-minimizing volumetric images, called OSD-CBCT, using surface images and low-frequency single-angle cone beam X-ray projections during radiation delivery. A-SI enables accurate, real-time tumor tracking throughout the radiation process, supporting precise radiotherapy. To implement this, we developed a patient-specific PC-DDPM, incorporating two key refinement strategies for generating OSD-CBCT from surface images. This process uses patient-specific data from pre-treatment 4DCT, and leverages projections collected at varying frequencies to improve anatomical accuracy. We also introduce physics-based geometric knowledge to further enhance OSD-CBCT quality. Our results show that PC-DDPM generates OSD-CBCT with excellent image and clinical quality, proving A-SI’s potential for real-world radiation delivery in radiotherapy. 
	
	\section{Declaration of Competing Interest}
	The authors declare that they have no known competing financial interests or personal relationships that could have appeared to influence the work reported in this paper.
	
	\section{Acknowledgement}
	This research is supported in part by the National Institutes of Health under Award Number R01CA272991, R01DE033512, R56EB033332, R01EB032680, and P30CA008748.

	\noindent 
	
	\bibliographystyle{plainnat}  
	\bibliography{arxiv}      
	
\end{document}